# Design and simulation of a low dark current metal/silicon/metal integrated plasmonic detector

Elahe Rastegar Pashaki
Ph.D. student
Photonics Research Laboratory
Electrical Engineering Department
Amirkabir University of technology
Tehran, Iran
Email: e.rastegar@aut.ac.ir

Hassan Kaatuzian
Professor
Photonics Research Laboratory
Electrical Engineering Department
Amirkabir University of technology
Tehran, Iran
Email: hsnkato@aut.ac.ir

Abdolber Mallah Livani
Assistant Professor
Electrical Engineering Department
Mazandaran University of Science and Technology
Behshahr, Iran
Email: mallah_a@aut.ac.ir

*Abstract*— **Silicon-based waveguide plasmon detectors have a great research interest because of CMOS compatibility and integration capability with other plasmonic integrated circuits. In this paper, a balanced metal-semiconductor-metal (MSM) integrated plasmonic detector is proposed to isolate the output from dark current and made it suitable for low noise applications. Performance characteristics of the new device are numerically simulated. In a specific bias point (V = 3 V), the output current is estimated to be about 31.8 μA and responsivity is 0.1288 A/W for a device with 2 μm$^2$ area. Simulation results for this balanced Plasmon detector demonstrate considerable dark current reduction compared with unbalanced plasmon detectors. Our estimated theoretical I-V characteristic, fits appropriately with experimental curve results reported before.**

## I. Introduction

Recently, many pieces of research have been focused on detection of infrared (IR) wavelengths, in applications like optical communications, material inspection, imaging for thermal analysis and biomedical measurements [1, 2]. Detection of IR wavelengths based on electron-hole pair (EHP) generation mechanism suffers from considerable noise resulted by using low bandgap semiconductors. On the contrary, internal photoemission (IPE) mechanism makes sub-bandgap detection possible. Nevertheless, low detection responsivity of IPE process is a negative point, which can be compensated by creating a higher intensity of electric field, and as a result, increasing level of photon absorption and photo-generated carriers in plasmonic detectors, specifically in Plasmonic Integrated circuits [3, 4].

Plasmonic IPE photodetectors can be fabricated in parallel (waveguide) [5, 6] or vertical [7] source configurations. Silicon-based waveguide plasmon detectors are a key component in designing CMOS compatible integrated plasmonic circuits because of their capability with other plasmonic integrated circuits [8]. These detectors are composed of Schottky contacts and plasmonic waveguides. For instance, Metal-Semiconductor-Metal (MSM) waveguides can be used in IPE based Schottky plasmon detectors in symmetric or asymmetric configurations.

The sample reported in [6] is proposed as a novel class of asymmetric MSM-IPE detectors, which has, comparable performance parameters to state-of-the-art photodiodes while having a tiny footprint. Despite all these benefits, the proposed device has a large amount of dark current that is under the same enhancement process as detection current, which makes these devices inappropriate for low noise applications.

In this paper, a balanced structure [21] based on the reported detector of [6] has been proposed, in which the effect of dark current on the output load is considerably decreased, while appropriate responsivity and bandwidth characteristics of the new device are relatively preserved. In section 2, physical structure of the proposed detector will be introduced. Then, operation principles will be discussed in sections 3. Simulation method and results are presented in section 4. Finally, process of this work will be concluded in section 5.

## II. Device Structure

Physical structure of the proposed detector, which consists of two identical asymmetric MSM waveguides, is sketched in fig.1.

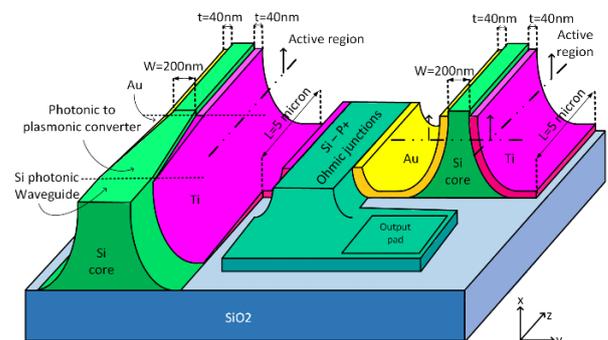

Fig. 1. Physical structure and dimensions of the proposed differential plasmon detector

Each waveguide has a lightly p-doped Silicon core sandwiched between Titanium and Gold layers forming two metal-semiconductor junctions. Physical parameters of the detector layers are summarized in table 1. In this structure, carriers flow through the narrower region of the Si core, which is considered as an active region (height ≈ 275nm) in fig. 1.

One of the two waveguides has an optical input while the other one is existed just for creating the same dark current as the main waveguide to form a differential structure, which leads to a considerable reduction of dark current in output.

TABLE I.    PHYSICAL PARAMETERS OF PROPOSED PLASMON DETECTOR

|  | Type /dope (cm-3) | Width (nm) | Length (µm) |
|---|---|---|---|
| Si-core | P-Type / $10^{15}$ | W = 200 | L = 5 |
| Si-P+ | P-Type / $10^{18}$ | 100 | L = 5 |
| Metals (Au, Ti) | -- | t = 40 | L = 5 |

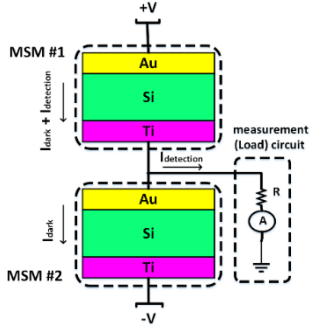

Fig. 2. Bias and measurement circuits of the differential detector.

MSM waveguides are biased to create same dark currents as shown in fig.2 that results in a negligible output current in absence of photon/plasmon source. By applying an optical input to MSM#1, detection current will be added to its background dark current. This extra current will be the output of the differential detector.

### III. OPERATION PRINCIPLES

#### A. A model for IPE

Internal photoemission mechanism is based on generation of hot carriers by absorption of photons (or plasmons) in metal side of a Schottky barrier. IPE can be described as a 3-step process [3]. 1) Generation of hot carriers by absorption of photons /plasmons in metal side, 2) transmission and scattering of hot carriers toward semiconductor interface, 3) Emission of hot carriers from the Schottky barrier and creating detection current. In order to overcome the Schottky barrier ($\phi_B$), hot carriers should be generated by absorbing Photons/plasmons with the energy (hυ; h is Planck's constant and υ is optical frequency) more than $\phi_B$ and less than bandgap of semiconductor ($\phi_B<h\upsilon<E_g$) because above the upper limit of this condition, Electron Hole Pairs (EHP) generation mechanism will be dominant. Details of photon absorption and hot carrier generation in different metals are described by ab-initio approach in [9,10] which provides density states of hot carriers in metal side. This density of states can be imported in Fowler-Northeim equation for calculating the transmission coefficient of a triangular potential (Schottky) barrier [11] and as a result, determining detection current. However, Berini [12] described this 3-step process through a semi-classical model, which is simpler and more practical in the simulation of photonic/plasmonic devices. According to this model, internal quantum efficiency of a photo-detector is calculated by [12]:

$$\eta_i = \frac{1}{2}\left(1-\sqrt{\frac{\Phi_B}{h\upsilon}}\right)^2 \quad (1)$$

On the other hand, there is another notation for ηi as [12]:

$$\eta_i = \frac{I_p/q}{S_{abs}/h\upsilon} \quad (2)$$

Where $I_p$ and q are photocurrent and elemental charge and $S_{abs}$ is absorbed optical power, which is converted to incident optical power by $S_{abs}=A\ S_{inc}$ that A determines the merit of photonic to the plasmonic converter. Therefore, photo-detection current in a single Schottky interface is obtained as follows:

$$I_p = q\frac{AS_{inc}}{2h\upsilon}\left(1-\sqrt{\frac{\Phi_B}{h\upsilon}}\right)^2 \quad (3)$$

Effects of tunneling and barrier lowering mechanisms will also be included in this simplified model in section B.

#### B. Energy band diagram

Energy band diagram of each MSM waveguide of proposed detector structure is shown in fig. 3(a). Schottky barriers for electrons on each side of the MSM waveguide can be calculated by $\phi_{Bn}$=W-X. Where X is the electron affinity of Si and W is metal's work function. Since a Schottky diode operation is based on majority carriers, in p-doped Si-core MSM waveguide, all calculations should be done on holes. The Schottky barrier for electrons can be converted to holes by $\phi_{Bp}=E_g-(W-X)$. However, surface states and Fermi level pinning effect influence these potential barriers and experimental data are more reliable than above equations. Consequently, schottky barrier heights for electrons at Au-Si and Ti-Si interfaces are considered as 0.82ev [13] and 0.62ev [14] respectively, which would be 0.3ev and 0.5ev for holes.

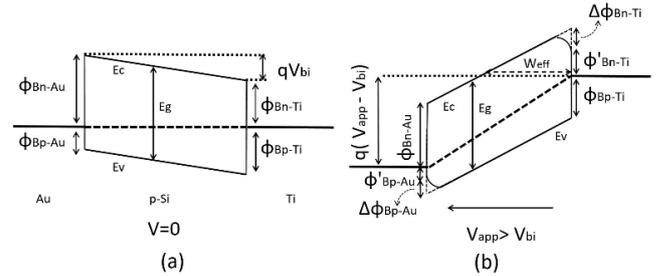

Fig. 3. Energy band diagram of Au-Si-Ti (a) without, (b) under bias.

In fig. 3(a), for compensation of internal voltage ($V_{bi}$=0.2V) arising from the Schottky barrier difference between two junctions, Au contact must be connected to a higher potential than the Ti side which leads to fig. 3(b). In this structure, surface plasmon polaritons (SPPs) propagate along both interfaces [15]. Plasmon absorption is proportional to imaginary part of metal permittivity at desired wavelength according to Beer's law ($\alpha=2\omega/c[\text{Im}(\varepsilon_r)/2]^{0.5}$ where α is the attenuation coefficient in metal, $\varepsilon_r$ and ω are metal's relative permittivity and angular frequency of photons and c is speed of light) [15]. The relative permittivity of gold and titanium can be calculated by the Drude-Lorentz model [16] as -93.06+11.11i and -4.87+33.7i at λ=1550nm respectively. According to these data, although the titanium side has a prominent role in plasmon absorption, generated hot holes of Au side determine detection current according to energy diagram of fig. 3(b) and this current will be added to the current of forward biased Ti-Si junction.

Under an appropriate biasing condition, energy diagram of fig. 3(a) changes into the fig. 3(b). Effects of Schottky barrier lowering ($\Delta\varphi_{Bn-Ti}$, $\Delta\varphi_{Bp-Au}$) and tunneling of carries through effective barrier widths ($W_{eff}$) should be considered in this case. The Schottky effect is the image-force-induced lowering of the potential energy for charge carrier emission and is proportional to applied voltage according to the following equation [17]:

$$\Delta\phi_B = \sqrt{\frac{qV_{app}}{4\pi\varepsilon_s W}} \qquad (4)$$

Here, $V_{app}$ is the applied voltage, $\varepsilon_s$ is Si permittivity and W is core width in the MSM structure. By reducing the Schottky barrier height, thermionic emission current will be increased.

Increasing the probability of tunneling through the Schottky barrier is another voltage-dependent effect in described MSM detector. Detailed energy diagram of Ti-(n-Si)-Au structure is shown in fig. 4.

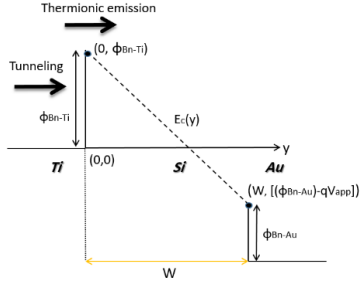

Fig. 4. Detailed energy diagram of Ti-Si-Au structure.

Accordingly, the relation between $V_{app}$ and conduction band energy ($E_c$) will be:

$$E_c(y) = \frac{\Phi_{Bn-Au} - \Phi_{Bn-Ti} - qV_{app}}{W} y + \Phi_{Bn-Ti} \qquad (5)$$

Tunneling component of current in a Schottky junction is shown in fig.4 and can be described by [18]:

$$J_T = \frac{A^*T}{K} \int_E^\infty \Gamma(E') \ln\left[\frac{1+fs(E')}{1+fm(E')}\right] dE' \qquad (6)$$

Where $J_T$ is tunneling current density, $A^*$ and $T$ are effective Richardson's coefficient and lattice temperature, $f_s(E)$ and $f_m(E)$ are the Maxwell-Boltzmann distribution functions in the semiconductor and metal, K is Boltzmann's constant, E is the carrier energy and $\Gamma(E)$ is tunneling probability. To obtain the localized tunneling rate, eq. (6) is imported in $G_T = (\nabla J_T)/q$ and yields [19]:

$$G_T = \frac{A^*T\vec{E}}{K} \Gamma(y) \ln\left[\frac{1+n/\gamma_n N_c}{1+\exp[-(E_c - E_{FM})/KT]}\right] \qquad (7)$$

Where E and n are the local electric field and local electron concentration, $N_c$ is the local conduction band density of states, $\gamma_n$ is the local Fermi-Dirac factor, $E_c$ is the local conduction band edge energy and $E_{FM}$ is the Fermi level in the contact. The tunneling probability ($\Gamma(y)$) in this equation can be determined by assuming a linear variation of conduction band energy ($E_c$), as follow [19]:

$$\Gamma(y) = \exp\left[\frac{-4\sqrt{2m}\,y}{3\hbar}\left(E_{FM} + \Phi_{Bn-Ti} - E_c(y)\right)^{\frac{1}{2}}\right] \qquad (8)$$

Here, m is the electron effective mass for tunneling and $\hbar$ is reduced Planck's constant. Based on Eq. (5), in a specific "y" by increasing the applied voltage, $E_c(y)$ will be decreased which causes enhancement of tunneling probability according to Eq. (8). Similar expressions of the above equations exist for holes by considering $E_v = E_c-E_g$, hole's average effective mass, $\phi_{BP}$, etc. as an alternative for corresponding parameters. However, calculation (based on parameters which are determined in Table. II of section IV) shows that effect of tunneling term in total current is very limited ($J_T$ is about $8.8435\times10^{-12}$), hence, it is neglected in later calculations.

*C. Current Analysis*

The differential detector of fig. 2 can be considered as 2 Schottky diodes in the configuration of fig. 5(a). Gold-Silicon (p-type) junction has a low potential barrier which reduces even more base on barrier lowering effect by applying the external voltage ($\Delta\phi_B = 0.0351$ev for V = 2V). Simulations in the next section show that Au-Si junction in this structure can be considered as an Ohmic contact and this assumption leads to an I-V characteristic that fits appropriately with experimental curve reported in [6].

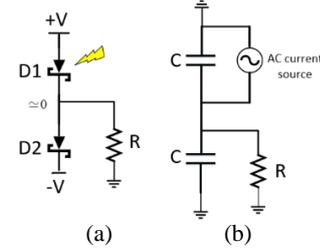

Fig. 5. (a) 2-diode and (b) capacitive model of differential plasmon detector

In this circuit, D1 and D2 are forward biased and hot holes generation occurs in Au-Si junction ($J_{det}$) which is imported in current relation of D1. Current density of these Schottky diodes can be determined by modifying relations of [17] as follow:

$$J_1 = \underbrace{(A^*T^2 e^{-\frac{\Phi_{Bp-Ti}}{KT}} + J_{det.})e^{\frac{qV_{D1}}{nKT}}}_{semiconductor \rightarrow metal} - \underbrace{A^*T^2 e^{-\frac{\Phi_{Bp-Ti}}{KT}}}_{metal \rightarrow semiconductor}$$

$$= (A^*T^2 e^{-\frac{\Phi_{Bp-Ti}}{KT}} + J_{det.})\left(e^{\frac{qV_{D1}}{nKT}} - 1\right) + J_{det.} \qquad (9)$$

$$J_2 = A^*T^2 e^{-\frac{\Phi_{Bp-Ti}}{KT}} \left( e^{\frac{qV_{D2}}{n_{dark}KT}} - 1 \right) \quad (10)$$

$$V_{D1} = V - area \times J_1 \times R_{Au-si} - \underbrace{area \times (J_1 - J_2) \times R}_{I_R}$$

$$V_{D2} = V - area \times J_2 \times R_{Au-si} + \underbrace{area \times (J_1 - J_2) \times R}_{I_R} \quad (11)$$

In these equations, $n_{dark}$ and $n_{det}$ are ideality factors of non-illuminated and illuminated modes respectively. In illuminated mode because of hot carrier generation in metal, average energy of electrons is more than dark condition that results in enhancement of thermionic emission process. Furthermore, Ideality factor in a Schottky barrier depends on contribution of thermionic emission and tunneling in diode current [22, 23]. Thus, $n_{dark}$ and $n_{det}$ are not equal and are specified as $n_{dark} = 1$ and $n_{det} = 3.5$ by curve fitting on experimental results of [6].

*D. Bandwidth*

Operating speed analysis of Au-Si-Ti detector has been done in [6]. It was shown that among different limitations, such as hot-carrier lifetime in metals, carrier drift time through the semiconductor layer and RC time constant, the last one has a dominant effect. Capacitance of MSM junction is estimated based on an equivalent parallel-plate capacitor (C) with 5μm × 275nm metal area across W=200 nm Si core which leads to a capacitance below C = 0.7 fF. According to fig. 5(b), in the differential detector, there would be two parallel capacitors, which duplicate RC time constant of this structure with same load resistance (R = 50 Ω) compared with the initial detector. Nevertheless, this bandwidth reduction (by factor 2) is the paid cost for eliminating the dark current effect on the load.

## IV. SIMULATION AND RESULTS

Simulations have been done in MATLAB™ to determine I-V characteristic of the structure shown in fig. 5(a). In order to verify the validity of mentioned equations (9, 10), the fabricated structure reported in [6], is simulated and results are compared with the reported empirical curves. Simulation parameters are summarized in table II and comparison of results are available in fig. 6 for dark and detection currents. Structural similarities between Introduced device of this work and reported device of [6], allow us to apply the presented theories to simulation of the proposed device.

TABLE II.  SIMULATION PARAMETERS

|        | Value | Ref. |              | Value | Ref. |
|--------|-------|------|--------------|-------|------|
| $A^*$  | 0.66×120 A/cm²/K² (p-type Si) | [17] | $\phi_{Bp-Ti}$ | 0.5 ev | [14] |
| $E_g$  | 1.12 ev @ room temp. |     | $\phi_{Bp-Au}$ | 0.3 ev | [13] |
| m      | 0.3×9.109×10⁻³¹ Kg (average mass of light and heavy holes) |     | A = $S_{abs}/S_{inc}$ | 0.5 |      |
| $N_v$  | 1.8×10¹⁹ for Si | [20] | $S_{inc}$ | 310 μW |      |
| $\gamma_p$ | -9.2 |     | T | 300 K |      |
| λ      | 1550 nm |     | $R_{load}$ | 50 Ω |      |

Numerical solution of equations (9) to (11) has been done simultaneously and I-V characteristics of different parts in the proposed differential detector are presented in fig. 7 and compared with the single MSM structure. Current of D1 which consists of dark and detection components is almost equal to the current of previous single MSM detector under illumination. On the other hand, D2 without any optical input has only the dark component, which is approximately equal to the dark current of the single MSM structure. However, there is a slight difference between corresponding currents in differential and single MSM detectors caused by increased load voltage. Load current in fig. 7 is created by the current difference of D1 and D2 in the differential structure and as mentioned before, it is equal to the detection current.

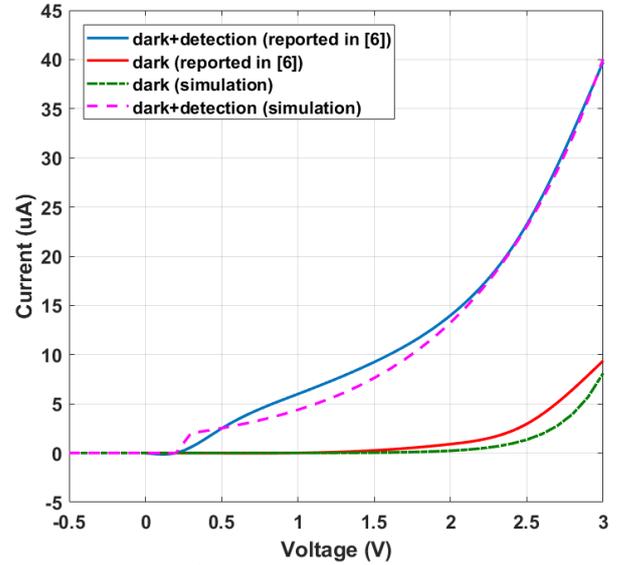

Fig. 6. Comparison of Simulation results with empirical curves [6]. Both experiment and simulation results belong to single MSM detector.

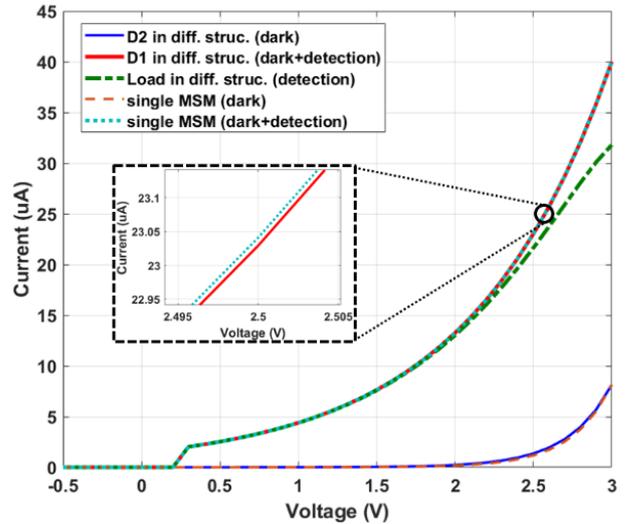

Fig. 7. I-V characteristics of different parts in the proposed differential detector and single MSM detector.

The responsivity of a detector is defined as the slope of output current versus optical input power characteristic, which is plotted for 3 bias voltages 1V, 2V and 3V in fig. 8.

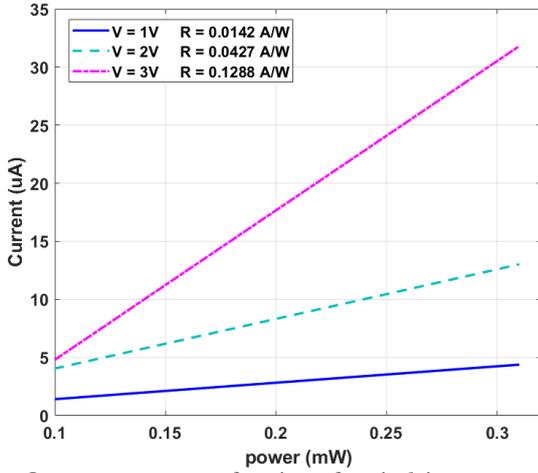

Fig. 8. Output current as a function of optical input power in the differential structure for V = 1, 2 and 3 V

Finally, In order to evaluate performance of the proposed device, dark current, responsivity, electrical bandwidth (BW) and area parameters are calculated for both structures and summarized in table III.

TABLE III. COMPARISON OF SIMULATED RESULTS IN DIFFERENTIAL (BALANCED) AND SINGLE MSM PLASMON DETECTORS (@V=3V)

|  | *Differential proposed Detector* | *Single MSM* |
|---|---|---|
| Dark Current | negligible | 8.11 μA |
| Photocurrent | 31.81 μA (load current) | 40.04 μA (dark + detection) |
| Respossivity | 0.1288 A/W | 0.1292 A/W |
| Electrical BW factor | x 0.5 | x 1 |
| Active region area (WxL) | ≈ 2 μm$^2$ | ≈ 1 μm$^2$ |

## V. CONCLUSION

In this paper, a differential asymmetric MSM IPE plasmon detector is proposed and theoretically analyzed. The key advantage of the proposed device is isolating the output from dark current while maintaining other performance characteristics in a reasonable range. However, this improvement result is in a trade-off with area and modulation bandwidth. Performance of the new device is theoretically investigated with semi-classic models. In a specific bias condition (V=3v), output current is estimated to be 31.8 μA, responsivity is predicted to be about 0.1288 A/W and bandwidth factor is x0.5. These properties, which are achieved in a small footprint 2 μm$^2$, make the new device a good choice for low noise integrated plasmonic applications.

REFERENCES

[1] R. Stanley, "Plasmonics in the mid-infrared," nature photonics, vol. 6, pp. 409-411, 2012.
[2] Mark L. Brongersma, Naomi J. Halas, and Peter Nordlander, "Plasmon-induced hot carrier science and technology," nature nanotechnology, vol. 10, pp. 25-34, January 2015.
[3] P. Berini, "Surface plasmon photodetectors and their applications," Laser Photonics Rev., DOI 10.1002/lpor.201300019, 2013.
[4] H. A. Atwater and A. Polman, "Plasmonics for improved photovoltaic devices," Nat. Mater. 9, 205–213, 2010.
[5] Goykhman, B. Desiatov, J. Khurgin, J. Shappir, U. Levy, "Waveguide-based compact silicon Schottky photodetector with enhanced responsivity in the telecom spectral band," Opt. Express 20, 28594– 28602, 2012.
[6] S. Muehlbrandt, A. Melikyan, T. Harter, and *et. al*. "Silicon-plasmonic internal-photoemission detector for 40 Gbit/s data reception," OPTICA Optical Society of America, Vol. 3, No. 7, pp. 741-747, 2016.
[7] S.R.J. Brueck, V. Diadiuk, T. Jones, W. Lenth, "Enhanced quantum efficiency internal photoemission detectors by grating coupling to surface plasma waves," Appl. Phys. Lett., vol. 46, pp. 915–917, 1985.
[8] A. M. Livani and H. Kaatuzian, "Design and simulation of an electrically pumped Schottky junction based plasmonic amplifier," Applied Optics, vol. 54, no. 9, pp. 2164−2173, 2015.
[9] Ravishankar Sundararaman, Prineha Narang, Adam S. Jermyn, William A. Goddard & Harry A. Atwater, "Theoretical predictions for hot-carrier generation from surface plasmon decay," nature communications, Dec. 2014.
[10] Marco Bernardi, Jamal Mustafa, Jeffrey B. Neaton & Steven G. Louie, "Theory and computation of hot carriers generated by surface plasmon polaritons in noble metals," nature communications, Jun. 2015.
[11] Richard g. Forbes and Jonathan h. b. Deane, "Transmission coefficients for the exact triangular barrier: an exact general analytical theory that can replace Fowler & Nordheim's 1928 theory," Proc. R. Soc. A vol. 467, pp. 2927–2947, 2011.
[12] Christine Scales, and Pierre Berini, "Thin-Film Schottky Barrier Photodetector Models," IEEE Journal of Quantum electronics, vol. 46, no. 5, pp. 633-643, May 2010.
[13] T. P. Chen, T. C. Lee, C. C. Ling, C. D. Beling, and S. Fung, "Current transport and its effect on the determination of the Schottky-barrier height in a typical system: gold on silicon," Solid-State Electron. 36, 949–954, 1993.
[14] M. Cowley, "Titanium-silicon Schottky barrier diodes," Solid-State Electron. 13, 403–414, (1970).
[15] S. A. Maier, Plasmonics: Fundamentals and Applications, Springer Verlag, 2007.
[16] B. Ung and Y. Sheng, "Interference of surface waves in a metallic nanoslit," Optics Express, 2007.
[17] S.M. Sze, Physics of semiconductor devices, Wiley-Interscience publication, chap. 5, 2nd ed., 1981.
[18] Matsuzawa, K., K. Uchida, and A. Nishiyama, "A Unified Simulation of Schottky and Ohmic Contacts", IEEE Trans. Electron Devices, Vol. 47, No. 1, pp. 103-108, Jan. 2000.
[19] Silvaco, Inc. (2013, Oct. 2). Atlas User's Manual [online]. Available: http://www.silvaco.com/
[20] Raseong Kim and Mark Lundstrom, "Notes on Fermi-Dirac Integrals," arXiv, 2008.
[21] A. Beling et al., "Monolithically integrated balanced photodetector and its application in OTDM 160 Gbit/s DPSK transmission," in Electronics Letters, vol. 39, no. 16, pp. 1204-1205, 7 Aug. 2003.
[22] Klyukanov, A & A. Gashin, P & Scurtu, R., "Ideality factor in transport theory of Schottky barrier diodes," arXiv: 1204.0335, 2012.
[23] H. Rhoderick and R. H. Williams, Metal-Semiconductor Contacts, Oxford University Press, 2nd ed., 1988.